\documentstyle[12pt,epsfig]{article}
\input psfig
\input psfig
 1
\def\as{\alpha_{\rm S}}

\def\citenum#1{{\def\@cite##1##2{##1}\cite{#1}}}
\def\citea#1{\@cite{#1}{}}

\def\as{\alpha_{\rm S}}

\def\b{\beta}
\def\a{\alpha}

\def\g{\gamma}

\def\l{\lambda}

\def\m{\mu}

\def\pr{\prime}

\def\s{\sigma}

\def\({\left(}
\def\){\right)}

\def\citenum#1{{\def\@cite##1##2{##1}\cite{#1}}}
\def\citea#1{\@cite{#1}{}}

\def\bt{b_{\perp}}

\def\bt2{$b^2_t$}

\def\jol1{$J_0(\,l_{1\perp}\,r_{\perp}\,)$}

\def\citea#1{\@cite{#1}{}}

\def\beq{\begin{equation}}
\def\eeq{\end{equation}}
\def\bea{\begin{eqnarray}}
\def\eea{\end{eqnarray}}

\def\eq#1{{eq.~(\ref{#1})}}

\def\bbbz{{\mathchoice {\hbox{$\sf\textstyle Z\kern-0.4em Z$}}
{\hbox{$\sf\textstyle Z\kern-0.4em Z$}}
{\hbox{$\sf\scriptstyle Z\kern-0.3em Z$}}
{\hbox{$\sf\scriptscriptstyle Z\kern-0.2em Z$}}}}
\def\npb#1#2#3{    {\it Nucl. Phys. }{\bf B#1} (19#2) #3}
\def\plb#1#2#3{    {\it Phys. Lett. }{\bf B#1} (19#2) #3}
\def\pr#1#2#3{     {\it Phys. Rev. }{\bf #1} (19#2) #3}
\def\prd#1#2#3{    {\it Phys. Rev. }{\bf D#1} (19#2) #3}
\def\prep#1#2#3{   {\it Phys. Rep. }{\bf #1} (19#2) #3}
\def\prl#1#2#3{    {\it Phys. Rev. Lett. }{\bf #1} (19#2) #3}

\def\zpc#1#2#3{    {\it Z. Phys. }{\bf C#1} (19#2) #3}

\def\sjnp#1#2#3{   {\it Sov. J. Nucl. Phys. }{\bf #1} (19#2) #3}

\relax

\begin{document}

\begin{titlepage}
\begin{flushright}
FERMILAB - PUB - 96 - 242 - T\\
CBPF - NF - 046/96 \\
{\bf hep - ph / 9608????}\\
August 26, 1996\\
\end{flushright}
\begin{center}
{\Large\bf Diffractive  Production  of  $ {\bf b {\bar b}}$ in  Proton -
Antiproton}\\ 
{\Large \bf Collision at the Tevatron }\\[12ex]
{\large G I L V A N\,\, A L V E S${}^{a)}{}^*$\footnotetext{${}^*$ E-mail:
gilvan@lafex.cbpf.br},\,\,\, E U G E N E \,\, L E V I N${}^{b)\,c){}^{\dagger}}$
\footnotetext{${}^{\dagger}$ E-mail: levin@fnal.gov;
leving@ccsg.tau.ac.il}}\\
 {\large and}\\
{\large   A L B E R T O \,\,
 S A N T O R O${}^{a)\,c){}^{\star}}$\footnotetext{${}^{\star}$ E-mail:
 santoro@fnal.gov;santoro@lafex.cbpf.br}}  
\\[4ex]
{$ {}^{a)}$\it  LAFEX, Centro Brasileiro de Pesquisas F\'\i sicas  (CNPq)}\\
{\it Rua Dr. Xavier Sigaud 150, 22290 - 180 Rio de Janeiro, RJ, BRAZIL}
\\[2.5ex]
{${}^{b)}$\it Theory Department, Petersburg Nuclear Physics Institute}\\
{\it 188350, Gatchina, St. Petersburg, RUSSIA}\\[2.5ex]
{${}^{c)}$\it Fermi National Accelerator Laboratory}\\
{\it P.O. 500 Batavia  Illinois -60510 U.S.A.}\\[9.5ex]
\end{center}
{\large \bf Abstract:}
We show that the cross section of the diffractive production of $b \bar b$
can be described as the sum of two contributions: the first is proportional to
the probability of finding a small size $b \bar b$ color dipole in the 
fast hadron wave function before the interaction with a target, whilst
the second is the $b \bar b$-production after or during the interaction
with the target. The formulae are presented as well as the discussion of the
interrelation between these two contributions and the Ingelman -Schlein and
coherent diffraction mechanisms. The main prediction is that the coherent
diffraction mechanism  dominates, at least at the Tevatron energies,
giving the unique possibility to study it experimentally. 

\end{titlepage}

\section{ Introduction. }
\begin{figure}[p] 
\centering
\setlength{\unitlength}{1.0in}
\begin{tabular}{c c}
\psfig{figure=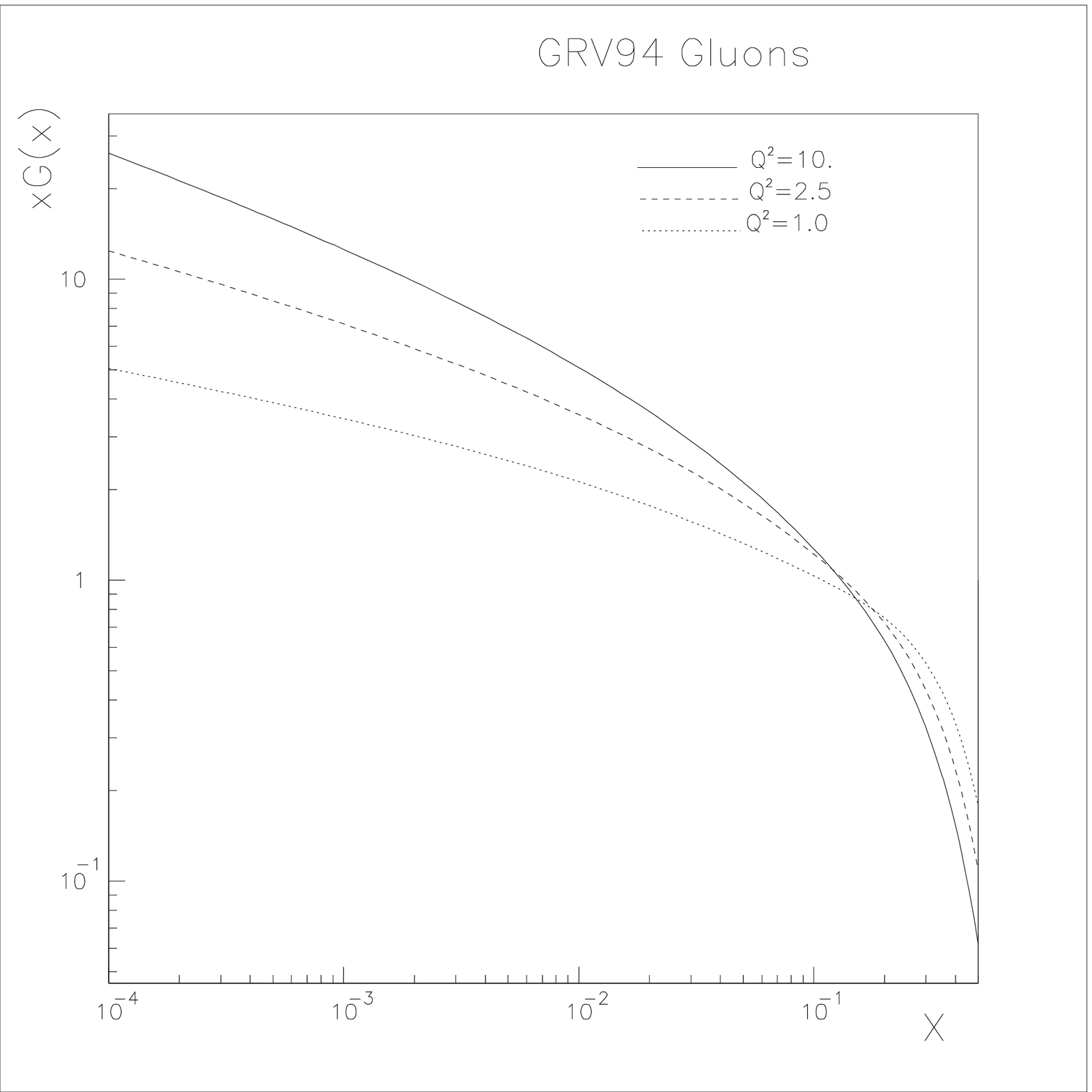,width=3.0in} & \psfig{figure=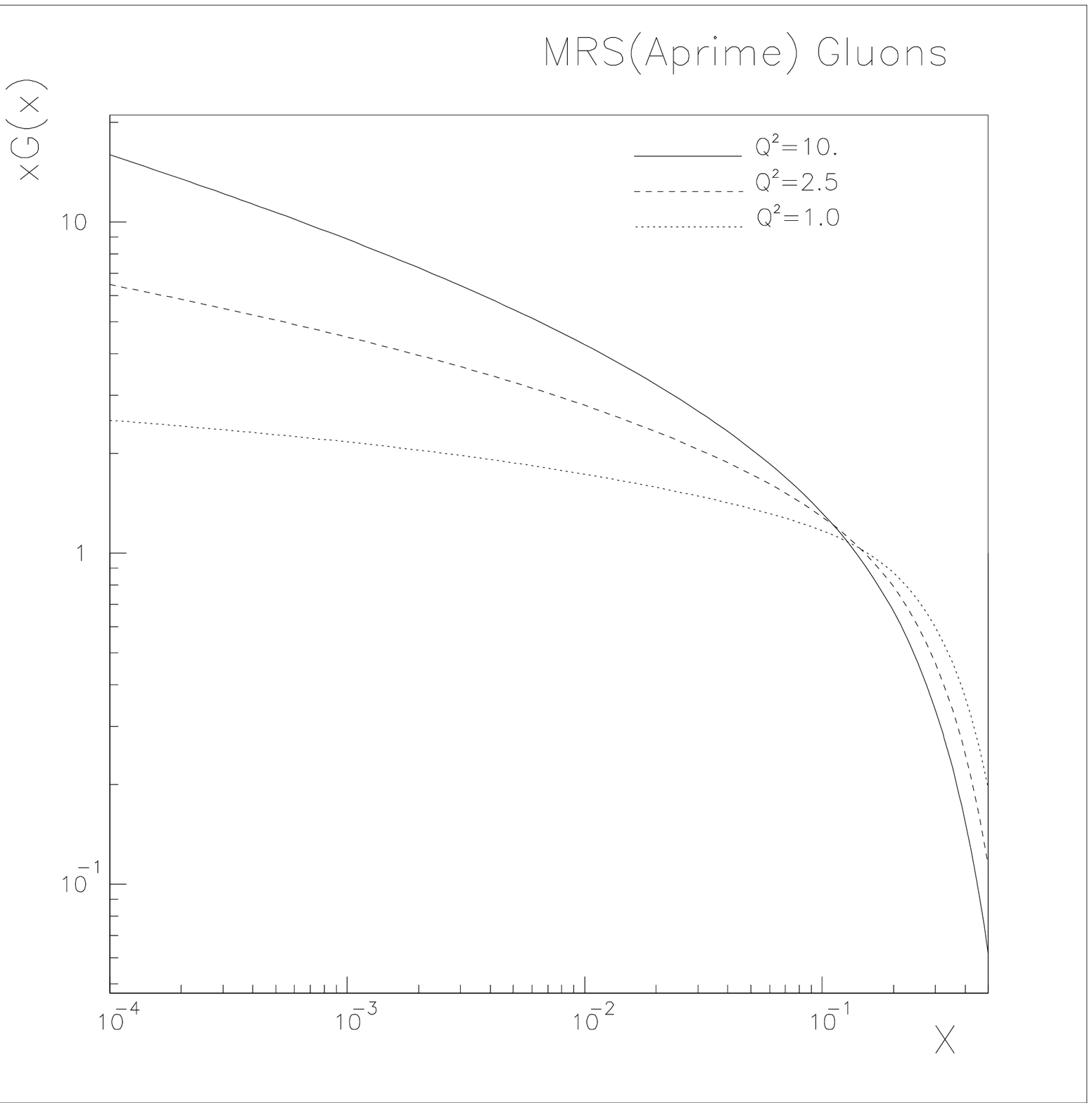,width=3.0in}\\ 
Fig.\ref{gluon}a & Fig.\ref{gluon}b\\
                 &                 \\
\multicolumn{2}{c}{\psfig{figure=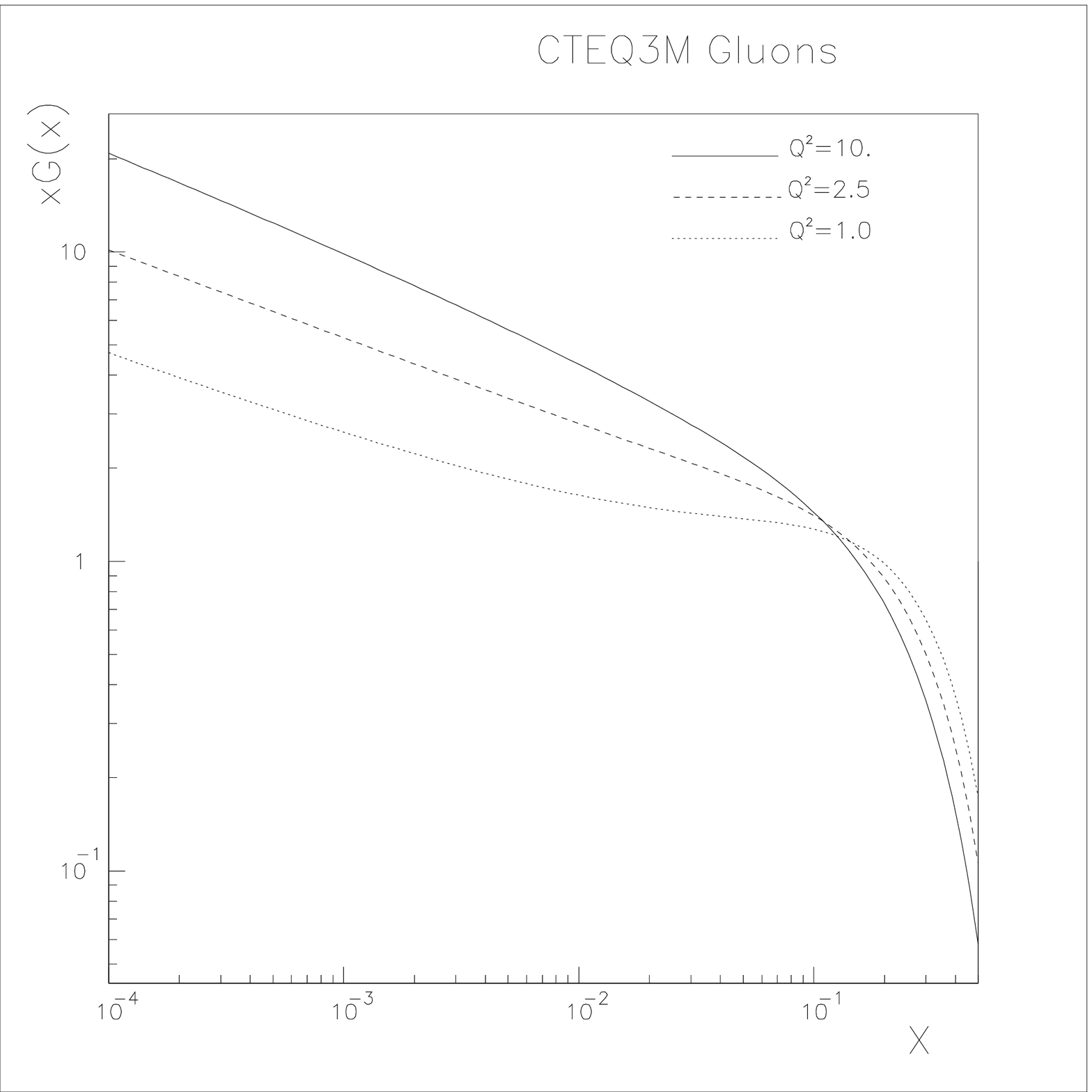,width=3.0in}}  \\
\multicolumn{2}{c}{Fig.\ref{gluon}c}  \\
\end{tabular}
\protect\caption{ Gluon structure function  $x G(x,Q^2)$ in different 
parameterizations: GRV [2] (~Fig.\ref{gluon}a ), 
MRS(A')[3] (Fig.\ref{gluon}b ) and CTEQ [4]  ( Fig.\ref{gluon}c).}
\protect\label{gluon}
\end{figure}

The main goal of this letter is to consider 
 the possibility of measureing the inelastic cross section in the diffractive
kinematic region and to discuss
the diffractive production of 
$ b \bar b $ - pair as a way to extract the value of gluon structure function
 $( x_{Bj} G(x_{Bj}, Q^2) )$ 
in the region of small $x_{Bj}$. New HERA data \cite{HERA}
shows a rapid increase of
$F_2(x_{Bj},Q^2)$ in the region of small $x_{Bj}$ ( $x_{Bj} \,<\,10^{-2} $ ),
which could be interpreted as  a 
manifestation of the growth of the gluon structure function
at small $x_{Bj}$. However, the data on $F_2$ does not allow the extraction 
of the value of the gluon structure function within good accuracy. At the 
present we know the gluon structure function  with accuracy up to factor two
(~see  Fig. \ref{gluon}, that shows the gluon 
structure in three  parametrizations GRV94 \cite{GRV94}, MRS(A) \cite{MRSA} 
and CTEQ \cite{CTEQ} at different values of $Q^2$
as function of $x_{Bj}$ ). Data on photoproduction of $J/\Psi$
seems to favor the MRS(A) parametrization \cite{JPROD}. This question, however,
is still open.

\begin{figure}[htbp]
\centerline{\psfig{figure=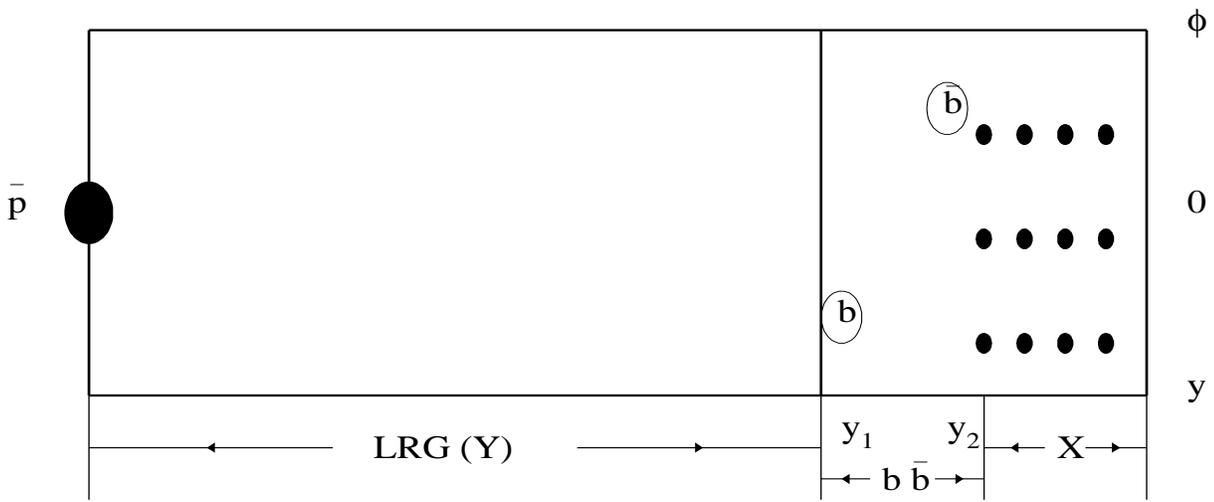,height= 12cm,width=20cm}}
\caption{Lego - plot of $b \bar b$ diffractive production in $p \bar p$
 collision.}
\protect\label{LEGO}
\end{figure}

\begin{figure}[htbp]
\centerline{\psfig{figure=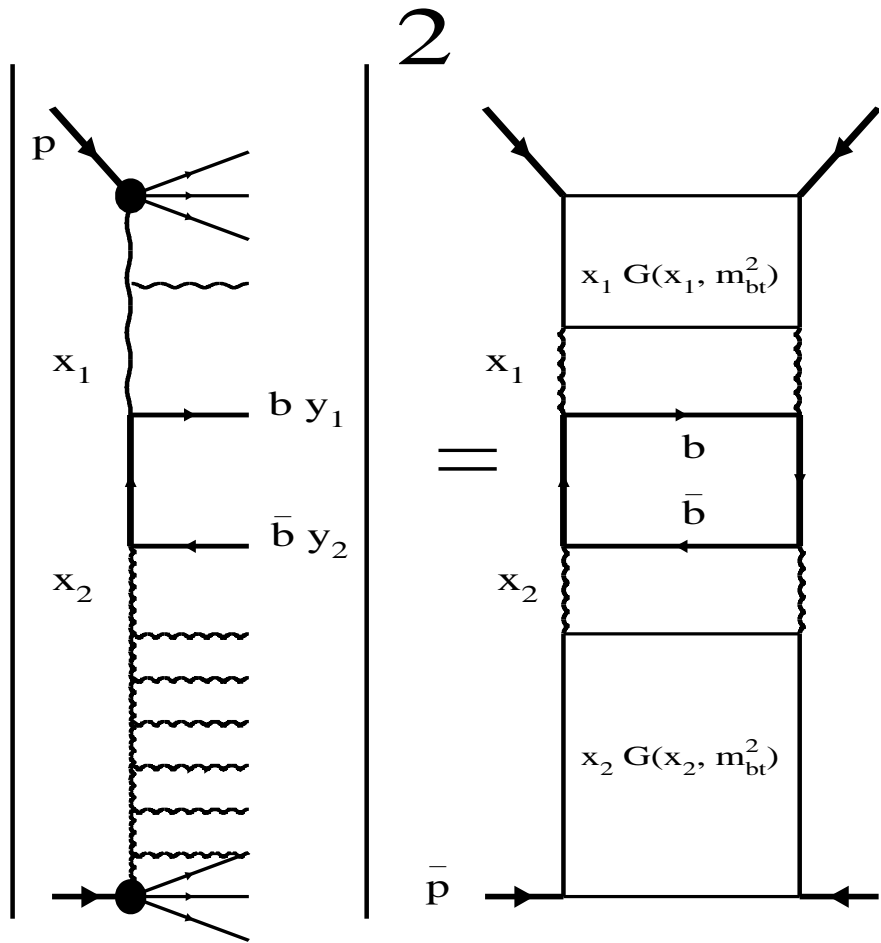,height= 17cm,width=17cm}}
\caption{Optical Theorem.}
\protect\label{OPTICAL}
\end{figure}

 We will argue that the large rapidity coverage collider detectors at the 
Tevatron  offer an  unique oportunity
to measure the gluon structure function  at $2\,GeV^2\,\le\, Q^2 \,\le\, m^2_b + p^2_t$,
where $m_b$ is the $b$ - quark mass and $p_t$ is its transverse momentum,
at $10^{-4}\,<\, x \,<\,10^{-2}$ using the process of the diffractive 
dissociation of proton into $b \bar b $ - pair. This process lego plot
and amplitude are pictured on Fig. \ref{LEGO} and Fig.\ref{AMPLITUDE}
respectively.
It is clear from these
figures that this process is a typical large rapidity gap (LRG) process, 
suggested by Bjorken \cite{BJ}. As pointed out by Bjorken, and as we 
demonstrate below, such a process can be described as the exchange
of a ``hard" Pomeron, which could be rewritten through the gluon structure
function due to the intimate relation between inelastic and elastic process 
given by the optical theorem (Fig.\ref{OPTICAL}), 
( see ref. \cite{FERMIWG} \cite{DPFWG}for more details).

 We will show  that the cross section of the diffraction dissociation
 ( DD) can be described as the sum of two different contributions 
\footnote{In what follows we  use at large  the parton picture of 
interaction. It is easier to discuss the diffractive processes in this 
picture in the frame where the antiproton  is at rest  ( the fixed target 
frame). Of course, all results  will be  given in relativistic 
invariant way.}:    

1.  the first 
is proportional to the probability of finding a $\bar b b$  color dipole
with a small size, of the order of $r^2_t \,\propto\,\frac{1}{m^2_b + p^2_t}$,
in the fast hadron wave function before the interaction with the
target. This dipole scatters with the target and produces the measured
 final state of the DD process. This mechanism has a normal partonic 
interpretation and, in the Bjorken frame for the projectile ( in other
words in the frame where the $\bar b b$ color dipole is at rest),
it looks as a measurement of the partonic content of the Pomeron and 
corresponds to the Ingelman-Schlein (IS) hypothesis of the Pomeron structure 
function \cite{IS}.

2. the second is the production of the $\bar b b$ - pair after or during
 the interaction with a target. We will show that this mechanism corresponds
 to the so called coherent diffraction ( CD )
(see ref.\cite{CFS}) and we will demonstrate that the measurement
of the $\bar b b$ diffraction will allow, thanks to  the the different 
dependance on the transverse momenta of produced quarks for both 
mechanisms, the separation of the CD contribution from the (IS) one. 

We would like to stress that the above two contributions are closely
related to the classic diffractive dissociation picture suggested by 
Good and Walker \cite{GW} 25 years ago. Indeed, there are two different  
possibilities for the dissociation of a hadron $h$ into a pair of 
hadrons ( $h_1 $ and $h_2$ ): first, the beam particle ($h$) interacts
with the target and  dissociates into  the pair of hadrons 
( $h_1$ and $h_2$ ); second, the beam particle dissociates first and 
one of the produced particle interacts with the target
 ( see ref. \cite{SANTORO} for details).

However, we will argue that the diffractive production of the heavy quark
system is originated from the small distances where we can develop a theoretical
approach based on perturbative QCD (pQCD). The pQCD approach allows us to 
calculate the diffractive dissociation process of $\bar b b $ in such details 
which are beyond  our reach in ``soft" high energy phenomenology.

\section{Notations and kinematics.}
1. $y \,=\,\frac{1}{2}\,\ln\frac{E \,+\,p_L}{E \,-\,p_L}$ is the rapidity of
a particle with energy $E$ and longitudinal momentum  (along the beam 
direction ) $p_L$. For the rapidity in the center of mass frame we use 
the notation $y^*$.

2. $P_1$ and $P_2$ are the momenta of colliding proton and antiproton 
(in the c.m. frame $P_1 = P_2$):

\beq \label{1}
P_1 \,=\,\lbrace \frac{\sqrt{s}}{2}\,(\,1 \,+\,\frac{2\,m^2}{s}\,), 
\frac{\sqrt{s}}{2}, 0 , 0 \,\rbrace\,\,;
\eeq
$$
P_2 \,=\,\lbrace \frac{\sqrt{s}}{2}\,(\,1 \,+\,\frac{2\,m^2}{s}\,), 
 -\,\frac{\sqrt{s}}{2}, 0 , 0 \,\rbrace\,\,.
$$
 
3. $y_1$ and $y_2$ are the rapidities of produced $b$  and $\bar b$ quarks, 
$p_{1t}$ and $p_{2t}$ are their transverse momenta and $m_b$ is the $b$ mass.

4. $M^2$ is the mass of the produced $b \,\bar b $ -pair. $m$ is the mass of 
the proton or antiproton. $s$ is the squared energy of the reaction in the 
c.m. frame and it is equal to $s \,=\, ( P_1 + P_2)^2$

5. $\Delta y \,=\, y_1 \,-\,y_2 $ is the difference of rapidities between 
the produced $b$ and $\bar b $.

6. $Y\,=\,\frac{ y_1\,+\,y_2}{2}$ is the mean rapitity of the $\bar b b$ 
system.

7. $m^2_{it}\,=\,m^2_b \,+\,p^2_{it}$ where $i$ = $1,2$.

8. For the purpose of obtaining the kinematic relation in the simplest way we
use the Sudakov decomposition \cite{SUD} of the momenta of all particles, namely
\beq \label{2}
p_{i \mu}\,\,=\,\a_i\,P_{1 \mu} \,+\,\b_i\,P_{2 \mu }\,+\,p_{it \mu}\,\,,
\eeq
where vector $\vec{p}_{it}$ is orthogonal to $P_{1\mu}$ and $P_{2\mu}$.

At high  energy 
$p^2_{i \mu}\,=\,\a_i\,\b_i\,s\,-\,p^2_{i t}$ and the rapidity of particle 
$``i"$ is equal to
\beq \label{3}
y^*_i \,=\,\frac{1}{2}\,\ln\frac{\a_i}{\b_i}\,\,.
\eeq

9. Using  Eqs.(1),(2) and (3) we obtain, for produced $b$ quarks:
\beq \label{4}
\a_1\,\,=\,\,\frac{m_{1 t}}{\sqrt{s}}\,\,e^{y^*_1}\,\,;\,\,\,\,\,\,\,
\b_1\,\,=\,\,\frac{m_{1 t}}{\sqrt{s}}\,\,e^{-\,y^*_1}\,\,;\,\,\,\,\,\,\,
\a_2\,\,=\,\,\frac{m_{2 t}}{\sqrt{s}}\,\,e^{y^*_2}\,\,;\,\,\,\,\,\,\,
\b_2\,\,=\,\,\frac{m_{2 t}}{\sqrt{s}}\,\,e^{-\,y^*_2}\,\,.
\eeq
and
\beq \label{5}
M^2\,\,=\,\,2\,m_{1 t}\,m_{2 t}\,cosh( \,\Delta y\, )
\,\,+\,\,
m^2_{1 t}\,\,+\,\,m^2_{2 t}
\eeq

\begin{figure}[htbp]
\centerline{\psfig{figure=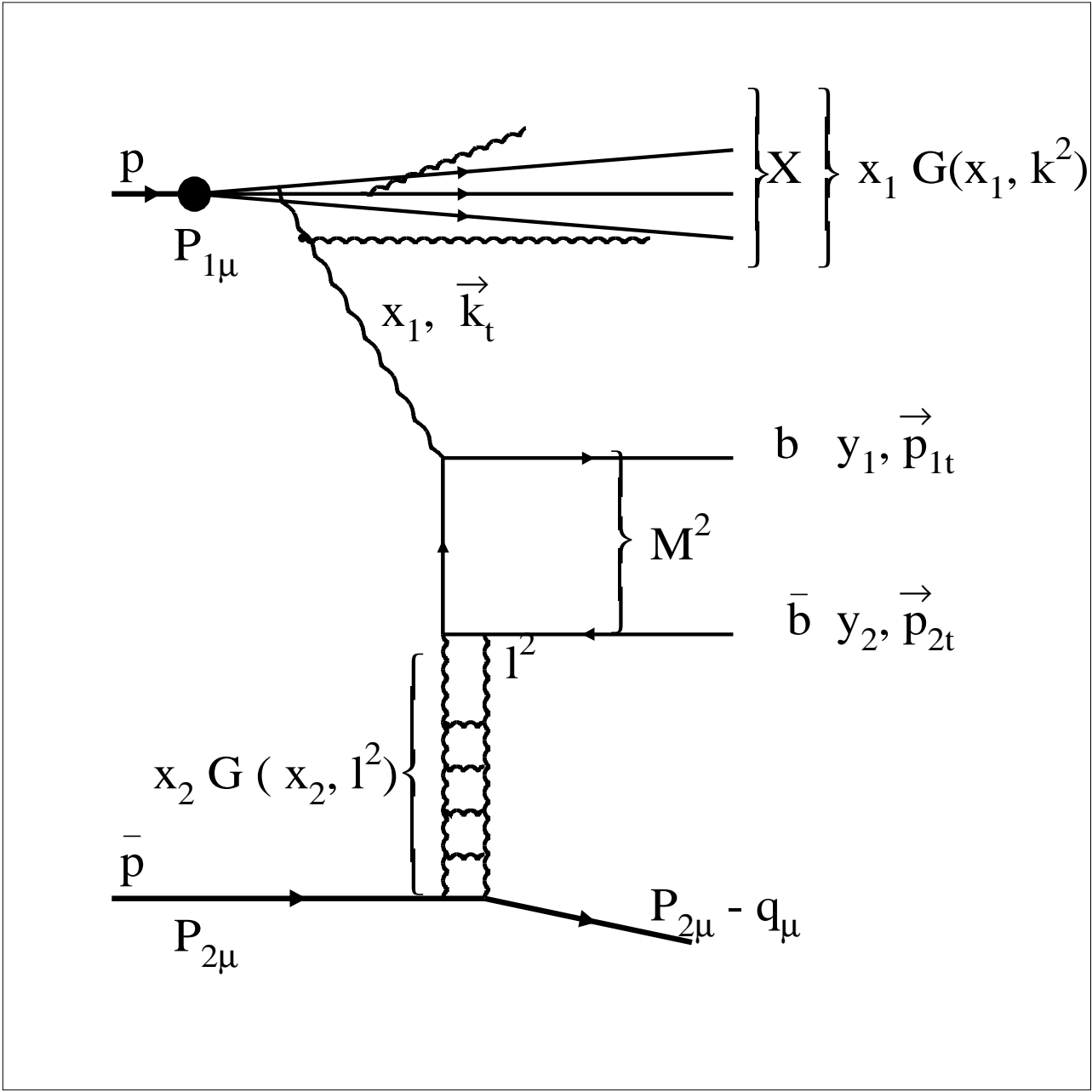,height= 17cm,width=17cm}}
\caption{Amplitude of $b \bar b$ - diffractive production.}
\protect\label{AMPLITUDE}
\end{figure}

10.  Let us introduce $x_1$ - the energy fraction  of hadron $``1"$ carried by
gluon $k$ in Fig.\ref{AMPLITUDE} and  - the energy fraction  of 
hadron $``2"$ carried by the Pomeron with momentum $q$ ( gluon ``ladder" 
in Fig.\ref{AMPLITUDE}).
We show below that $x_1$,$x_2$ will be the arguments of the gluon
structure functions in the cross section expression. Directly from
Fig.\ref{AMPLITUDE} one can see that
\beq \label{6}
x_1\,\,=\,\,\a_1\,\,+\,\,\a_2\,\,+\,\,\a_q\,\,;\,\,\,\,\,\,\,
x_2\,\,=\,\,\b_1\,\,+\,\,\b_2\,\,+\,\,\b_k\,\,,
\eeq
where $(\, x_1, \b_k\,)$ and $(\,\a_q, x_2\,)$ are the longitudinal components 
of the four momenta of gluon $1$ and the Pomeron, respectively.

The main property of high energy scattering is the fact that $\a_q
\,\ll\,\,\a_1\,\,and\,\,or\,\, \a_2$ and $\b_k\,\,\ll\,\,\b_1\,\,and\,\,or
\,\,\b_2$\,(see, for example, ref. \cite{GLR} ). Therefore, we can easily
derive from \eq{6}, assuming $m_{1 t}\,=\,m_{2 t}$:
\beq \label{7}
x_1\,\,=\,\,\frac{2\,m_{1 t}}{\sqrt{s}}\,e^{Y^*}\,cosh(\,\frac{\Delta y}{2}\,)
\,\,;\,\,\,\,\,\,\,x_2\,\,=\,\,\frac{2\,m_{1 t}}{\sqrt{s}}\,e^{-\,Y^*}\,
cosh(\,\frac{\Delta y}{2}\,)\,\,.
\eeq

11. Throughout the paper
 we will choose a frame where antiproton (see Fig.\ref{LEGO} and Fig.\ref{AMPLITUDE})
is essentially at rest and where all momenta ($l_i $) of fast particles
look as follows:

\beq \label{7b}
l_i\,=\,( l_{i \,+},l_{i\, -},\vec{l}_{i t} )\,=\,
\(\,l_{i\,+},\frac{m^2 + l^2_{t}}
{l_{i\, +},},
\vec{l}_t\,\)\,\,,
\eeq
where $l_{i +}\,=\,l_{i0} \,+\,l_{i 3}$ and $ l_{i -}\,=\,l_{i 0} \,-\,l_{i3}$.

12.  $x G(x,Q^2)$ everywhere in the paper is the gluon structure function.
 
\section{The value of the cross section in the generalized parton model.}

From Figures \ref{AMPLITUDE} and \ref{OPTICAL} we can see that the value of the 
cross section of our process
\beq \label{8}
p (P_1)\,\,+\,\,\bar p (P_2 )\,\,\rightarrow \,\,b (y_1, p_{1 t})\,\,+
\bar b ( y_2, p_{2 t } )\,\,+\,\,X\,\,+\,\,[ LRG ( Y )]\,\,+\,\,\bar p (P_2 - q)
\eeq
is equal to 
\beq \label{9}
\frac{d \s}{d Y d q^2_t d \Delta y d p^2_t}\,|_{q^2_t = 0}
\,\,=\,\,(x_1 \,G( x_1, \mu^2 )\,)\,\,\frac{d \s ^G
}{d Y d q^2_t d \Delta y d p^2_t}\,|_{q^2_t = 0}\,\,,
\eeq
where $\s^G$ is the reaction cross section.
\beq \label{10}
G(x_1, k^2_t)\,\,+\,\,\bar p (P_2 )\,\,\rightarrow \,\,b (y_1, p_{1 t})\,\,+
\bar b ( y_2, p_{2 t })\,\,+\,\,[ LRG ( Y )]\,\,+\,\,\bar p (P_2 - q)
\eeq
The physical meaning of \eq{9} is very simple: $x_1 G( x_1, \mu^2)$ is the
probability of finding a gluon with the fraction of energy $x_1$ inside
of the proton and $\s^G$ is the cross section of its interaction with 
the antiproton. In the spirit of the factorization theorem \cite{COLLINS} we
introduce the factorization scale $\mu^2$, the maximal value of $k^2_t$ at which we still can 
neglect the dependence of $\s^G$ on $k^2_t$. 

To simplify the color algebra we adopt throughout the paper the 
colorless probe approach, replacing the gluon with the transverse
momentum $k_t$ and the fraction of energy $x_1$ by a colorless probe
with the same kinematics.  The physical motivation is clear and based on the
 factorization equation  (\eq{9}). Indeed, we can measure the gluon structure
function using a colorless probe like the graviton or heavy Higgs boson.
The properties of such a probe have been studied in details in ref. \cite{MUELLER}.

The cross section for  the reaction of \eq{10} can be easily calculated.
It is clear that we have two mechanisms for $\bar b b$-production by
 the colorless probe which we will discuss in the rest target rest frame 
 ( antiproton in Fig.\ref{LEGO}).

 1. The first mechanism is the following:  there is a
$\bar b b $ component in the wave function of the fast probe
 before its interaction with the target. This $\bar b b$ pair is a color dipole
with sufficiently small
transverse size of the order of $r^2_t \,\propto\,\frac{1}{m^2_b + p^2_t}$
 which scatters with the target producing the measured final state.

2. The second mechanism is the production of the $\bar b b$ - pair after
 or during the interaction with the target.

\begin{figure}[htbp]
\centerline{\psfig{figure=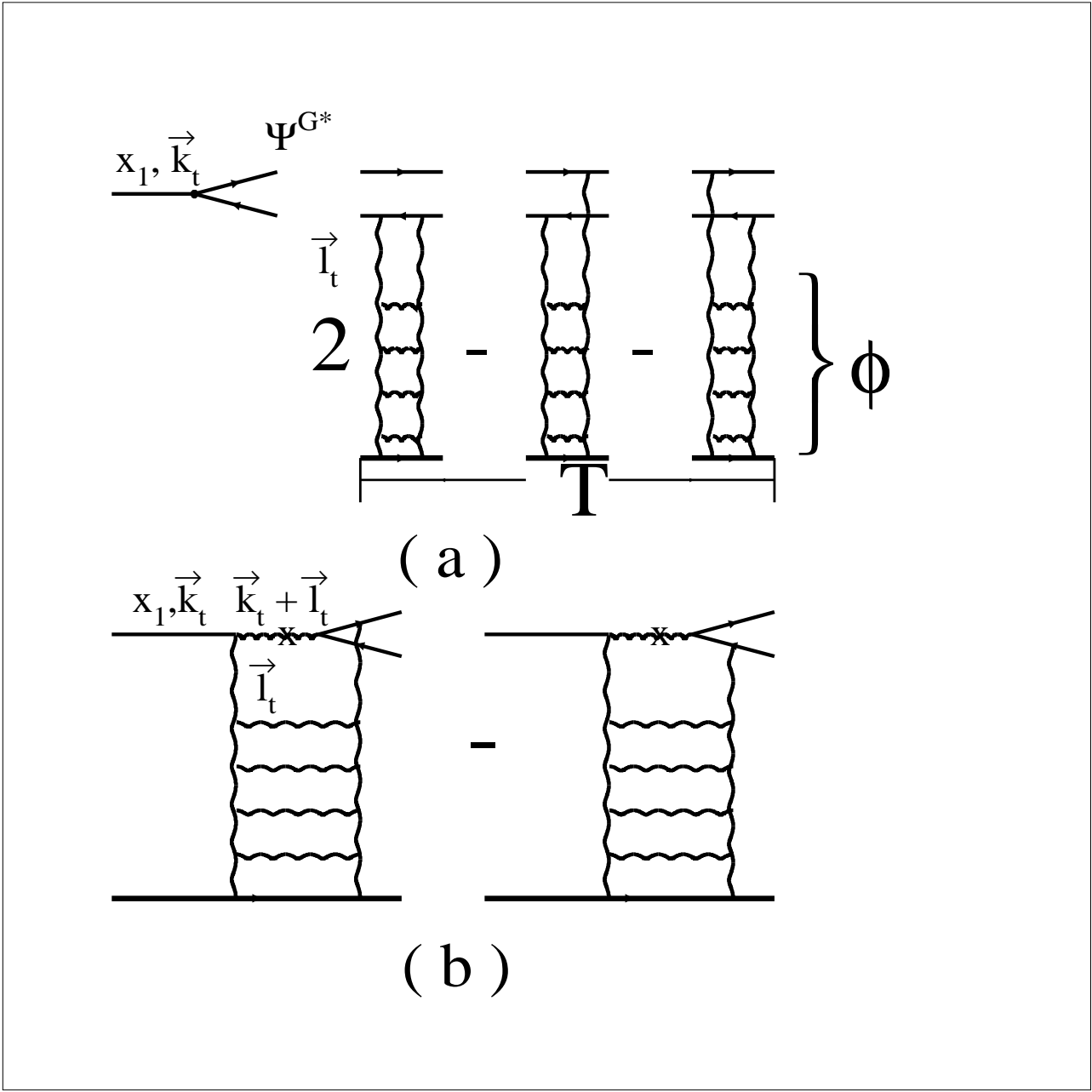,height= 17cm,width=17cm}}
\caption{Feyman diagrams for $b \bar b $ diffractive production by 
colorless gluon probe.}
\protect\label{FEYNMAN}
\end{figure}

These two mechanisms correspond to the two set of the Feyman diagrams pictured
 in figuress \ref{FEYNMAN}a and \ref{FEYNMAN}b, respectively.

 Let us start from the first one which looks normal from partonic point of
view in the sense that, in the Bjorken frame for the probe, it looks like the
 measurement of the partonic content of the Pomeron and corresponds to the
Ingelman - Schlein hypothesis of the Pomeron structure function \cite{IS}.
For the set of the Feyman diagrams of Fig. \ref{FEYNMAN}a 
 the amplitude of $\bar b b$ production can be written as  a product of
 two factors: (i) the wave function of $\bar b b$ pair in a virtual gluon
$\Psi^{G^*}_{\l_1 \l_2}$
 and (ii) the rescattering amplitude  of the quark - antiquark pair on the
 target $T_{\l_1 \l_2}$, where $\l_i$ is the quark polarization.
Following the conventions of ref.\cite{BRODKSKY}, we have:
\beq \label{11}
M_f\,\,=\,\,\sqrt{N_c} \,\int \,\,\frac{d^2 p'_t}{ 16 \pi^3} \,\int^1_0 d z'
\Psi^{G^*}_{\l_1 \l_2} (p'_t, z') \,\,T_{\l_1 \l_2} (p'_t, z'\,;\,p_t, z )\,\,,
\eeq 
where $p_t$ is the transverse momentum of the produced quark and $z$
 is the fraction of energy carried by $b$-quark with respect to
energy of the gluon. It is  easyly found from \eq{4} and \eq{6} that
\beq \label{12}
z\,\,=\,\,\frac{\a_1}{x_1}\,\,=\,
\,\frac{\a_1}{\a_1\,+\,\a_2\,+\,\a_q}\,\,=\,\,
\frac{e^{\frac{\Delta y}{2}}}{e^{\frac{\Delta y}{2}}\,\,+\,\,
e^{-\,\frac{\Delta y}{2}} }\,\,,
\eeq
where $\a_q$ we can be found from the equation: $(\, P_2 \,-\,q\,)^2 = m^2 $
and it is equal to $ \a_q\,=\,\frac{ q^2}{ (1 - x_2) s}\,\,\ll\,\a_1 + \a_2$ at
large $s$. In deriving \eq{12} we have also assumed that $\vec{p}_{1t}\,=\,
-\,\vec{p}_{2t} \,+\,\vec{k}_{t}\,+\,\vec{q}_t\,\rightarrow\, - \,\vec{p}_{2t}$.

The virtual gluon breaks into a
 quark - antiquark pair with a large lifetime which is equal to
$\tau_{G^*}$. In
leading log(1/x) approximation of pQCD, which we will use here, the time of
 interaction is much smaller than $\tau_{G^*}$ and during this time
the exchange of gluons does not change the fraction of energy carried by
quark or/and antiquark. It is instructive to recall the argument of why this 
is so.  According to the uncertainty principle the lifetime of the $\bar b b $
 fluctuation ($\tau_{G^*}$) is
\beq \label{t1}
\tau_{G^*}\,\,\sim\,\frac{1}{\Delta E}\,=\,\vert\, \frac{1}{ k_{-}\,-\,
p_{1\,-} \,-\,p_{2\,-}}\,\vert\,=\,\frac{x_1 \,z (1-z)\,P_{1 \,+}}{m^2_t\,
+\,z ( 1 - z) k^2_t}\,\,.
\eeq
An estimate of the interaction time can be obtained from the typical time  for
the emission of a gluon with momentum $l$, from the quark $p_1$, say. Then
\beq \label{t2}
\tau_{i}\,\,\sim\,\,\vert \frac{1}{p'_{1\,-} \,-\,p_{1\,-} \,-\,l_{-}} \vert
\,=\,\vert \frac{x_1 P_{1\,+}}{\frac{m^2_t}{z'}\,-\,\frac{m^2_t}{z} \,-\,
\frac{l^2_t}{ \a_l}} \vert\,\,,
\eeq
where $\a_l\,=\,\frac{l_{+}}{x_1\,P_{1\,+}}$ and $ z'\,=\,z\,+\,\a_l$. In the
 leading $\,log (1/ x)$ approximation of pQCD we have $\a_l\,\,\ll\,\,z$ 
and hence
\beq \label{t3}
\tau_i\,\approx\,\frac{\a_l x_1 P_{1\,+}}{l^2_t}\,\,\ll\,\,\tau_{G^*}\,\,.
\eeq
Therefore, the interaction only  changes the transverse momenta of 
quarks (see Fig.\ref{AMPLITUDE}). The vertices also do not depend on the type of the diagram
since the exchange of gluons preserves helicity at high energy.
Finally the amplitude $T$ can be reduced to the form \cite{BRODKSKY}:
\beq \label{13}
T_{\l_1 \l_2}\,\,=
\eeq
$$
=16 \pi^3 \int \{\, 2 \delta( \vec{k'}_t - 
\vec{k}_t)\,-\,\delta( \vec{k'}_t - \vec{k}_t - \vec{l}_t) \,-\,
\delta( \vec{k'}_t - \vec{k}_t + \vec{l}_t)\,\}\cdot\delta ( z - z')
\,\,\phi(l_t,x)\,\frac{d^2 l_t d l_{+} }{16 \pi^3 \,l^4_t}\,\,,\\ \nonumber
$$
where the function $\phi$ corresponds to the ``ladder" diagram (see Fig.\ref{AMPLITUDE}) 
and only weakly (logarithmically ) depends on $l_t$. $l_{+} $
 is the large component of  vector $l_{\mu}$ which we have introduced in the
previous section. The difference in signs between the terms in 
Eq.(17) reflects the different color charge of quark and antiquark.

 Substituting \eq{13}
in \eq{11} we obtain:
\beq \label{14} 
M_f\,\,=\,\,\sqrt{N_c} \,\int\,\,
\Delta\Psi^{G^*}_{\l_1 \l_2} (p_t,l_t, z) \,\,
\phi(l_t,x)\,\frac{d^2 l_t d l_{+} }{16 \pi^3 \,l^4_t}
\eeq
where 
\beq \label{15}
\Delta \Psi^{G^*}(p_t,l_t,z)\,=\,2\,\Psi^{G^*}(p_t,z)\,-\,
\Psi^{G^*}(p_t - l_t,z)\,-\,\Psi^{G^*}(p_t +l_t,z)\,\,.
\eeq

Function $\Psi$ has been found to be (see for example ref. \cite{BRODKSKY} ):
\beq \label{16}
\Psi^{G^*}_{\pm}(p_t,z)\,=\,- \,g\,\,
\frac{\bar{ u}_{\l_1}(p_1)\vec{ \g}\cdot 
\vec{\epsilon}^{G^*}
 v_{\l_2}(p_2)}{\sqrt{z \,(\,1 \,-\, z\,)} \,(\, k^2\,-
\,\frac{m^2_b + p^2_t}{z (1 - z)
}\,)}\,\,=
\eeq
$$
= -\,g \,\cdot\,\frac{1}{a^2 + p^2_t}\,\,\{
\delta_{\l_1\,-\l_2}
\,[\, \l_1 \,(1 - 2z) \,\pm\,1
\,]\,\vec{\epsilon}^{G^*}_{\pm}\,\cdot\,
\vec{p}_t\,\,\,\pm\,\,\,m_b\,\,
\delta_{\l_1 \,\l_2}\,]\,\}
$$
where  $\as \,=\,\frac{g^2}{4 \pi}$ ; $\vec{\epsilon}^{G^*}_{\pm} $
 is the circular polarization vector of the gluon
 \mbox { ( $ \vec{\epsilon}^{G^*}_{\pm}\,=\,\frac{1}{\sqrt{2}}\,
(\,0,1, \pm\,1,0\,)$)}
   and
 $ a^2 \,=\,m^2_b \,+\,k^2 z (1 - z )$. We have used formulae from 
 refs.\cite{LEPAGE} and \cite{MUELLER} in the above calculations.

Considering $l^2_t \,\ll\, m^2_{b t} $ we obtain :
\beq \label{17}
\Delta \Psi^{G^*}_{\pm}(p_t,l_t,z)\,\,=\,- 2\,g\,\, \cdot\,l^2_t \cdot
\eeq
$$
\cdot \{\,4\,\frac{a^2}{
(\,a^2 \,+\,p^2_t\,)^3}\,\,\delta_{\l_1\,\l_2}\,[\,\l_1( 1 - 2 z )\,\pm\,1\,
\,]\,\vec{\epsilon}^{G^*}_{\pm}\,\cdot\,\vec{p}_t\,\,+
\,\,\,m_b\,
\frac{a^2 \,-\,p^2_t}{
(\,a^2 \,+\,p^2_t\,)^3}\,\,\pm
\,\delta_{\l_1\,-\l_2}\,\}\,\,. 
$$

In the leading log approximation in ln$(1/x)$ and ln$(a^2/\Lambda^2)$
\cite{LEVIN,BRODKSKY} 
\beq \label{18}
\int  \phi(l_t,x)\,\frac{d^2 l_t d l_{+} }{16 \pi^3 \,l^2_t}\,\,=\,\,
i\,\frac{4\pi^2 T_R \as}{N_c} \,(s + k^2) x G(x,a^2 + p^2_t)\,\,,
\eeq
where $T_R/N_c$ arises from averaging  over colors $(T_R = 1/2)$.

Collecting all previous equations we can calculate cross section:
\beq \label{19}
\frac{d \s ^G}{d Y d q^2_t d \Delta y d p^2_t}\,|_{q^2_t = 0}\,\,=\,\,
\frac{\sum_{\l_1 \l_2} M^2_f}{16 \pi s^2 }\,\,\frac{d z}{d \Delta y}=\,\,
\frac{d z}{d \Delta y}\, \alpha \as^2 \frac{16 \pi^2}{9} \cdot
\eeq
$$
\cdot \,\{\,
\,[\, (z^2 + (1 - z )^2)\,p^2_t \,] \,(
 \,\frac{a^2}{(\,a^2\,+\,p^2_t\,)^3}\,)^2\,\,+
\,\,\frac{1}{4}\,
\,m^2_b\,(\,\frac{a^2\,-\,p^2_t}
{(\,a^2\,+\,p^2_t)^3}\,)^2\,\}\,\cdot \,(\,x G(x, a^2 + p^2_t))^2
$$

Finally,
we can rewrite \eq{19} in the form ( $N_c= 3 $ ):
\beq \label{20}
\frac{d \s ^G}{d Y d q^2_t d \Delta y d p^2_t}\,|_{q^2_t = 0}\,\,=
\eeq
$$\,\,
= \frac{ 16\, \pi^2\as^3}{9}\,\frac{1}{4\, cosh^2(\frac{\Delta y}{2})}\,
[\,
\frac{cosh( \Delta y )}{2 cosh^2(\frac{\Delta y}{2} )}p^2_t
\,\ \,\,+\,\,\frac{m^2_b}{4}\,
 ( \,1\,-\,\frac{p^2_t}{a^2}\,)^2\,]
\,\,\cdot\,\{\frac{a^2 }{( a^2 + p^2_t)^3}\}^2\,\,( x G(x, a^2 +  p^2_t)
 )^2\,\,.
$$

For the cross section of the diffractive dissociation we have
 (after sum over gluon polarization and correct averaging over color
 ($N_c$ = 3)):
\beq \label{21}
\frac{d \s}{d Y d q^2_t d \Delta y d p^2_t}\,|_{q^2_t = 0}
\,\,=\,\,\(\,x_1 \,G( x_1, \mu^2 )\,\)\,\,\cdot
\eeq
$$
\cdot \frac{16\, \pi^2 \as^3}{9}\,\frac{1}{4\,cosh^2(\,\frac{\Delta y}{2})}\,
[\,
\frac{cosh( \Delta y )}{2 cosh^2(\frac{\Delta y}{2} )}\,p^2_t \,\,+\,\,
\frac{m^2_b}{4}\,
(\,1 \,-\,\frac{p^2_t}{a^2}\,)^2]
\,\,\cdot\,\{\frac{a^2 }{( a^2 + p^2_t)^3}\}^2\,\,( x_2 G(x_2,
 a^2 + p^2_t ) )^2\,
$$

\begin{figure}[htbp]
\centerline{\psfig{figure=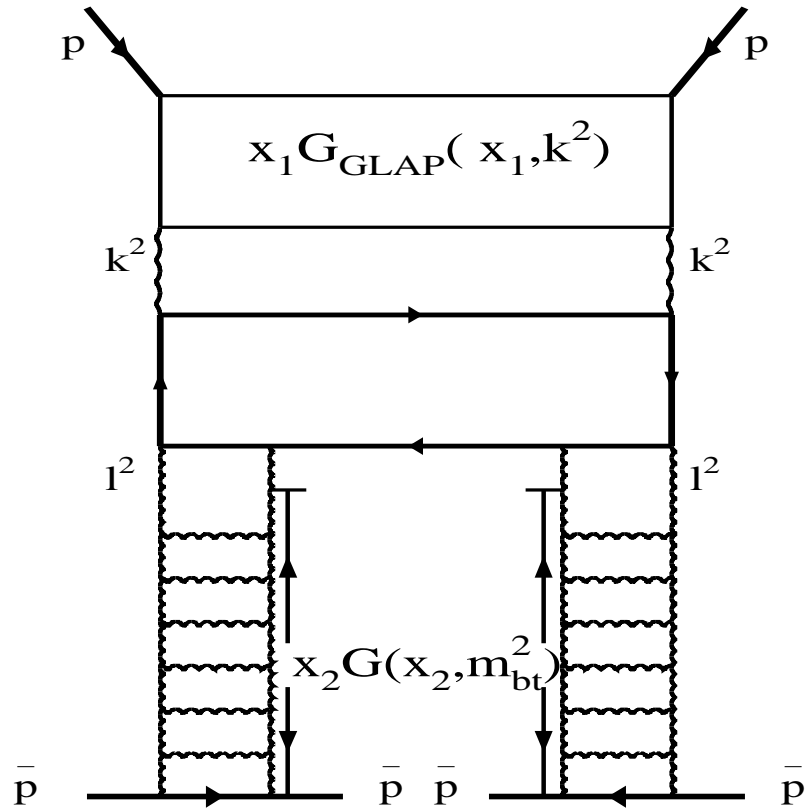,height= 17cm,width=17cm}}
\caption{The cross section for $ b \bar b $ diffractive production 
in the Ingelman - Schlein approach
[9].}
\protect\label{LADDER}
\end{figure}

From the  expression for $a$ we can set the factorization scale, since our
cross section ceases to depend on $k^2_t$ if $k^2_t\,\leq\,4 m^2_{bt}$.
Therefore, the reasonable choice is $\mu^2 = 4 m^2_{b t}$. We can neglect
the scale dependance in our cross section and put $a^2 =m^2_b$. All
ingredients of Eq.(25) are clearly seen in Fig. \ref{LADDER}.

	Taking into account the running QCD coupling constant we have to
	replace $\as^3$ in Eq.(25) by $ \as(\m^2)\as^2(a^2+p^2_t)$.

Now, let us consider the diagrams of Fig. \ref{FEYNMAN}b.
They correspond to the possibility of producing a  $\bar b b$ pair inside of the
 Pomeron. Indeed, the Pomeron is not  a point-like particle; 
gluons inside it live
  sufficiently long time  and during this time they
can create a $\bar b b$-pair which rescatters with the proton by exchange
 only one gluon. Indeed, the lifetime of $G_l G_l$ - pair in the diagram of 
Fig. \ref{FEYNMAN}b is equal to
 $\tau_l\,=\,\frac{x_1 \,s}{k^2_t \,+\,\frac{l^2_t}{z_l (1 - z_l)}}$, where 
$z_l$ is the energy fraction of gluon $l$. This time is much bigger than
the time $ \tau_b$ that $\bar b b$ - pair lives ($\tau_b \,=\,
\frac{x_1 \,s}{M^2} \,\ll\,\tau_l$).

As has been discussed many times (see, for example refs. \cite{MUELLER}
 \cite{KURAEV} \cite{LW})
we can safely calculate the diagram of fig. \ref{FEYNMAN}b
by closing the contour of integration over
$\b_l$ on the propagator marked by cross in Fig. \ref{FEYNMAN}b. 
We anticipate that  $l_t < k_t$ and that the vertex of emission of 
the gluon $1'$ is proportional to $l_{\mu t}$ (see ref.\cite{GLR}).
The interaction of the gluon with transverse momentum $k_t + l_t$
with the target is calculated  using \eq{14} with 
\beq \label{23}
\Delta \Psi(p_t,l_t,z)\,\,=\,\,\Psi(p_t + l_t,z)\,\,-\,\,\Psi(p_t - l_t,z)\,\,.
\eeq
Substituting \eq{16} in \eq{23} one obtains after integration over the
azimuthal angle of vector $l_t$:
\beq \label{24}
 \Delta \Psi(p_t,l_t,z)\,\,=\,\,- 2\,g\,\,\,l^2_t\,\vec{p}_t\cdot
\eeq
$$
\cdot\{\,\delta_{\l_1 \,-\l_2}\,[\, (1 - 2z)\l_1 \,\pm\,1\,]\,
\frac{a^2}{(\,a^2\,+\,p^2_t\,)^2}\,\,-\,\,\delta_{\l_1\,\l_2}\,\,
m_b\,\l_1\,
\,\frac{1}{(\,a^2\,+\,p^2_t\,)^2}\,\}\,.
$$  
Using \eq{18} and \eq{19} we obtain ($N_c$ = 3 ):
\beq \label{26}
\frac{d \s ^{G^*} [ CD]}{d Y d q^2_t d \Delta y d p^2_t}\,|_{q^2_t = 0}\,\,=
\eeq
$$\,\,
=\,\frac{4\, \pi^2\as^3}{9}\,\frac{1}{4\,cosh^2(\frac{\Delta y}{2})}\,
[\,\frac{cosh(\Delta y )}{2\,cosh^2(\frac{\Delta y}{2})}
\,a^4 \,\,+\,\,m^2_b\,p^2_t\,]\,\cdot\frac{1}{k^2_t}
\,\,\cdot\,\{\frac{1}{( a^2 + p^2_t)^2}\}^2\,\,( x_2 G(x_2, k^2_t)
 )^2\,\,.
$$
Notice that extra factor $1/k^2_t$ in \eq{26} comes from the fact that 
$\Delta \Psi$ of \eq{24} does not depend on the polarization of the gluon
with transverse momentum $k_t$ which is proportional to $k_t$ and cancels
one of the gluon propagators in \eq{19}. We would like to recall that in 
the previous calculation we assumed that $l_t \,<\,k_t$ and this
inequality establish the scale in the gluon structure function in \eq{26}.
The answer for the cross section of the coherent diffraction has the form:
\beq \label{27}
\frac{d \s [ CD]}{d Y d q^2_t d \Delta y d p^2_t}\,|_{q^2_t = 0}\,\,=
\int^{m^2_{bt}}\,\,\frac{d k^2_t}{k^4_t} \,\,\frac{\partial x_1 G(x_1, k^2_t)}{
\partial \ln k^2_t}\,\,\(\, x_2 G(x_2, k^2_t)\,\)^2\cdot
\eeq
$$\,\,
\cdot\frac{4\, \pi^2\as^3(k^2_t)}{9}\,\frac{1}{4\,cosh^2(\frac{\Delta y}{2})}\,
\,
\,\,\cdot\,[\,\frac{cosh(\Delta y )}{2\,cosh^2(\frac{\Delta y}{2})}
\,a^4 \,\,+\,\,m^2_b\,p^2_t\,]\,\cdot\,\frac{1}{(\,a^2\,+\,p^2_t\,)^4}\,\,.  
$$
This equation gives the contribution for so called coherent diffraction (CD)
 \cite{CFS}. The most contribution to the integral comes from the region of  
 sufficiently small $k_t$ due to the factor
$k^4_t$ in the dominator and for the proton $k_t \,\propto \,\frac{1}{R_p}$
where $R_p$ is the proton radius. It means that we cannot trust our
 perturbative calculation  for the CD contribution. However,if we calculate
the integral
\beq \label{28} 
\int^{m^2_{bt}}\,\,\frac{d k^2_t}{k^4_t} \,\,\frac{\partial x_1 G(x_1,
k^2_t)}{
\partial \ln k^2_t}\,\,\(\, x_2 G(x_2, k^2_t)\,\)^2\,\,,
\eeq
using the current parametrization of the gluon structure function, we can 
find out that the typical $k^2_t$, which is essential in the integral, is 
not very small but about 1 - 2 $GeV^2$. To understand this fact we have to 
remember that the gluon structure function behaves as $( k^2_t)^{<\g>}$ 
(at least in semiclassical approach) and the value of $<\g>$ calculated in 
the current parametrization for the gluon structure function turns out to be 
rather big in the region of $k^2 \,\approx \,1 - 2 GeV^2 $
 (see Fig. \ref{CSECMRS} ). One can see that if $<\g>\,>\,0.5$ the integral 
converges on the upper limit or, in other words, the small distances  start
to be important.

\begin{figure}[htbp]
\centering
\setlength{\unitlength}{1.0in}
\begin{tabular} {c c }
\psfig{figure=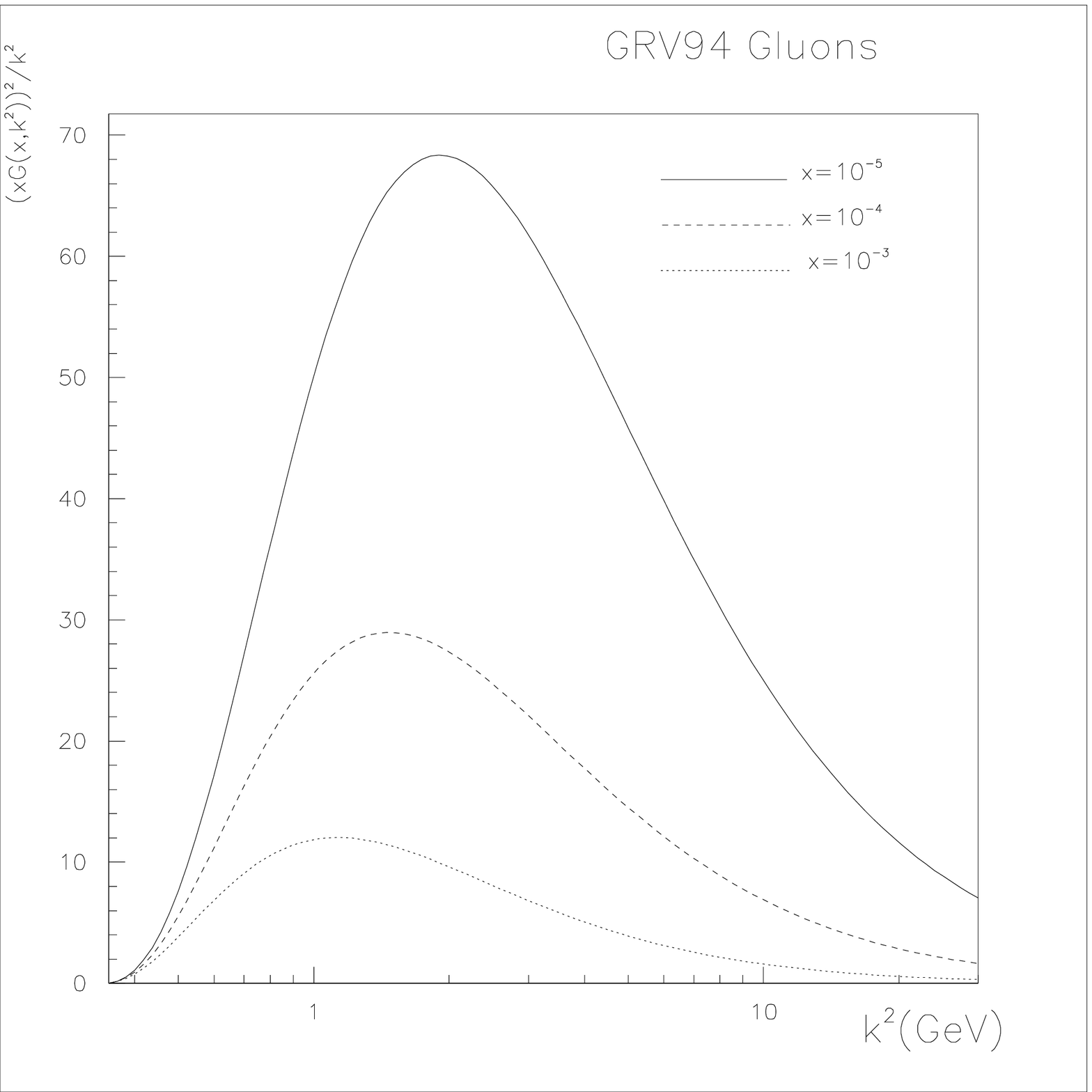,width=3.0in} & \psfig{figure=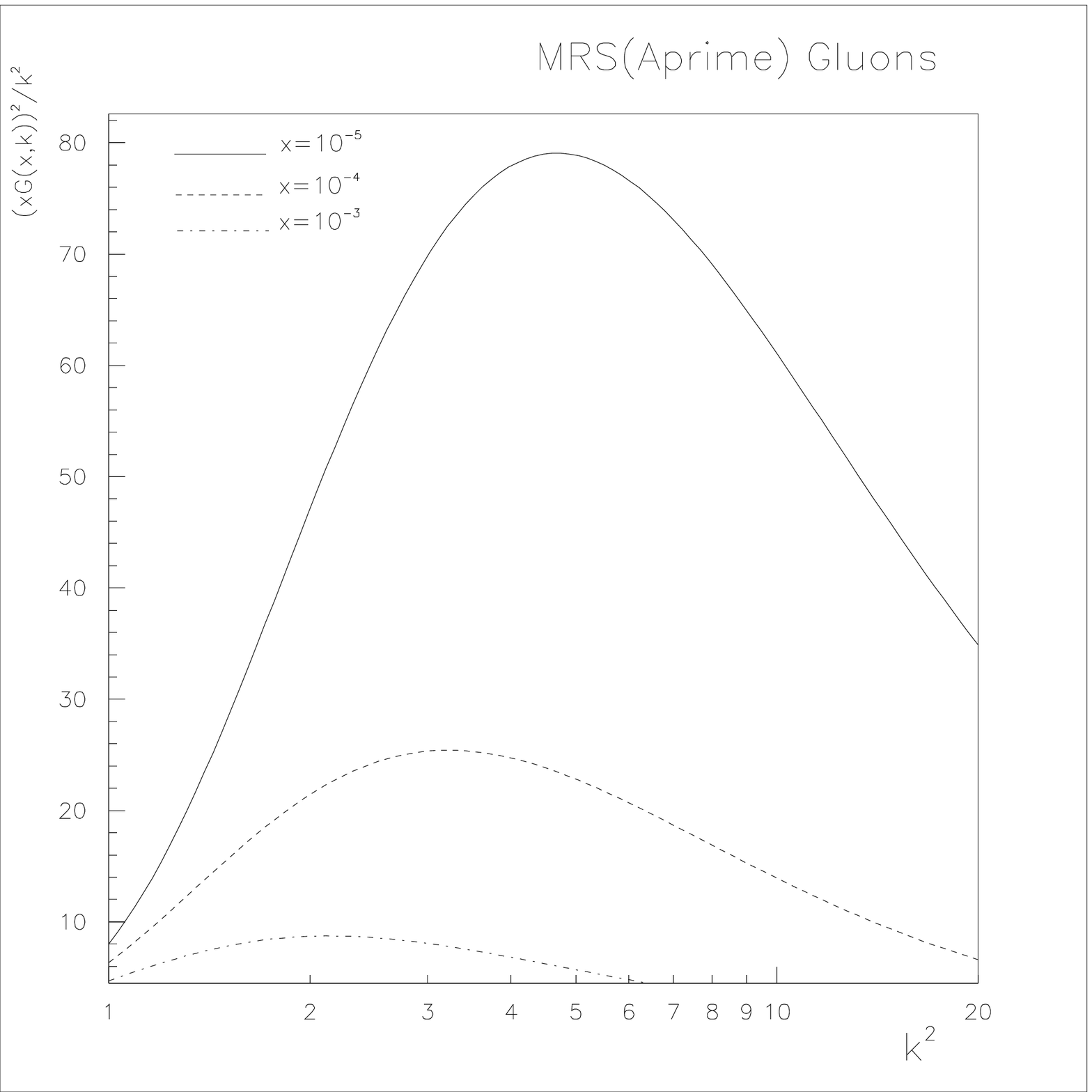,width=3.0in}\\
Fig. \ref{INTEGRANT}a  &  Fig.\ref{INTEGRANT}b \\
                       &                         \\
\multicolumn{2}{c}{\psfig{figure=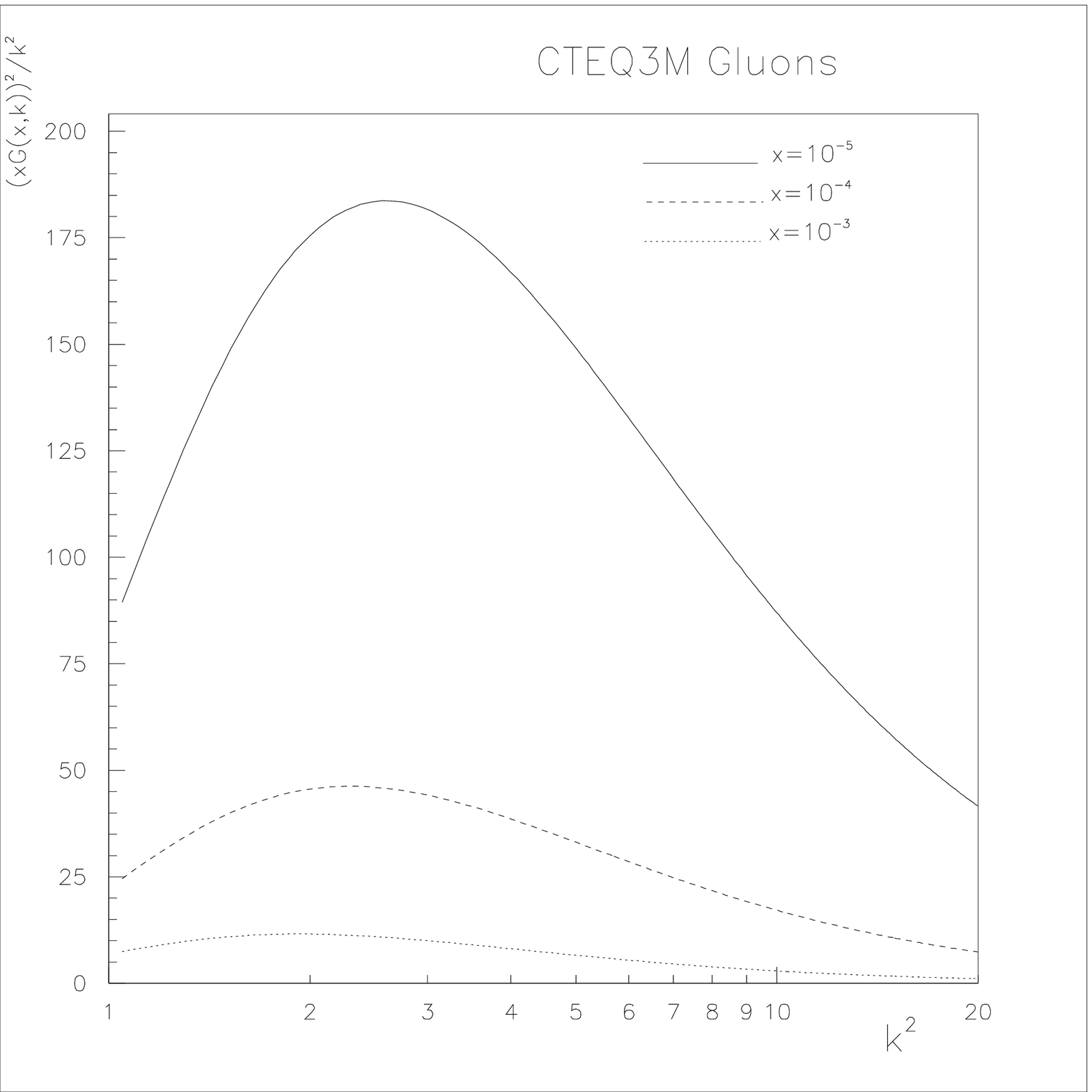,width=3.0in}}  \\
\multicolumn{2}{c}{Fig.\ref{INTEGRANT}c} \\
\end{tabular}
\protect\caption{ $ \frac{(x_2 G( x_2, k^2_t))^2}{k^2_t}$ versus 
$\ln( k^2_t/Q^2_0)$ in different parameterizations: GRV [2] ( Fig.1a ), 
MRS(A') [3] ( Fig. 1b ) and CTEQ [4] (Fig. 1c).} 
\protect\label{INTEGRANT}
\end{figure}

To check this statement we calculate the integrand of \eq{28} as a function of
$\ln ( k^2_t/Q^2_0)$ with  
$\frac{\partial x_1 G(x_1, k^2_t)}{\partial \ln k^2_t} \,=\,1$. $Q^2_0$
 = 0.34 $GeV^2$ is the
initial virtuality in the GRV parametrization. The result of the calculation
 is plotted in Fig. \ref{INTEGRANT}a for the GRV, in Fig. \ref{INTEGRANT}b for the 
MRS(A') and in Fig. \ref{INTEGRANT}c for
 the CTEQ parametrizations. We see a
 definite maximum in $\ln k^2_t/Q^2_0$ dependence  around $k^2_t \approx 1 - 1.5
 GeV^2$ which becomes more pronounced at smaller values of $x_2$. It means
 that we can safely use the perturbative QCD approach to calculate the CD 
contribution.

In numerical esstimates of  \eq{27} we use the GLAP equation \cite{GLAP}
to calculate
$\frac{\partial x_1 G( x_1, k^2_t)}{ \partial \ln k^2_t}$, namely :
\beq \label{GLAP}
\frac{\partial x_1 G( x_1, k^2_t)}{ \partial \ln k^2_t}
\,\,=\,\,\frac{\as(k^2_t)}{2\,\pi}\,\,\{\,\frac{4}{3} \,\int^1_x \,
\frac{d z}{z}\,[ z^2 \,+\,2 ( 1 - z)\,]\,\sum_i\,
\frac{x}{z}\,q_i(\frac{x}{z},k^2_t) \,\,+
\eeq
$$
\,+\,\, 6\,\int^1_x\,\frac {d z}{z}\,
[\,z^2( 1 - z) \,+\,1\,-\,z\,]\,\frac{x}{z}\,G( \frac{x}{z},k^2_t)\,\,+\,\,
6\,\int^1_x\frac{z d z}{ 1 - z}\,[ \,\frac{x}{z} G( \frac{x}{z}, k^2_t)
\,-\,x G(x k^2_t)\,]\,\,+
$$
$$
 +\,\,6\,[\,\frac{11}{12}\,-\,\frac{N_f}{18}\,]
\,x G(x,k^2_t)\,\,\}\,\,,
$$
where $N_f$ is the number of flavours and $N_C =3$ is the number of colors.
The running coupling QCD constant $\as(k^2_t)\,=\,\frac{ 4 \,\pi}{
(\,11\,-\,\frac{2}{3}\,N_f\,)\,\ln\frac{k^2_t}{\Lambda^2}}$.

It is worthwhile mentioning that in the case of the diffractive dissociation in 
the deep inelastic scattering the smallest value of $k_t $ is $k^2_t = Q^2$
and \eq{27} gives the contribution of the order of $1/Q^2$. In other
words, the coherent diffraction in this case is a high twist contribution while
the (IS) diffraction (see \eq{21}) occurs in  the leading twist.This
result has been obtained in ref. \cite{DDDIS}.

It should be stressed that there is no interference between the 
Ingelman-Schlein and the coherent diffraction contributions. Indeed,
the interference term vanishes due to integration over the azimuthal
angle of $p_t$ and summation over gluon polarizations, as one can see
comparing Eq.(21) and Eq.(27)

\section{Numerical estimates.}
Setting $x_1 =0.1$ we can estimate $x_2\,\approx\,0.6 \,10^{-3}$.
As far as the value of the cross section is 
 concerned, we get at $\Delta y = 0$
and $p_t = 0 $ the value ( at $\as$ = 0.25)
$$
\frac{d \s}{d Y d q^2_t d \Delta y d p^2_t}\,|_{q^2_t = 0}\,\,\approx\,\,
0.1 \,10^{-3} \frac{mbarn}{GeV^4}
$$
Here, we took $x_2G(x_2,m^2_{b t})$ = 20, which is 
the value in the GRV parametrization.

\begin{figure}[htbp]
\centering
\setlength{\unitlength}{1.0in}
\begin{tabular} {c c }
\psfig{figure=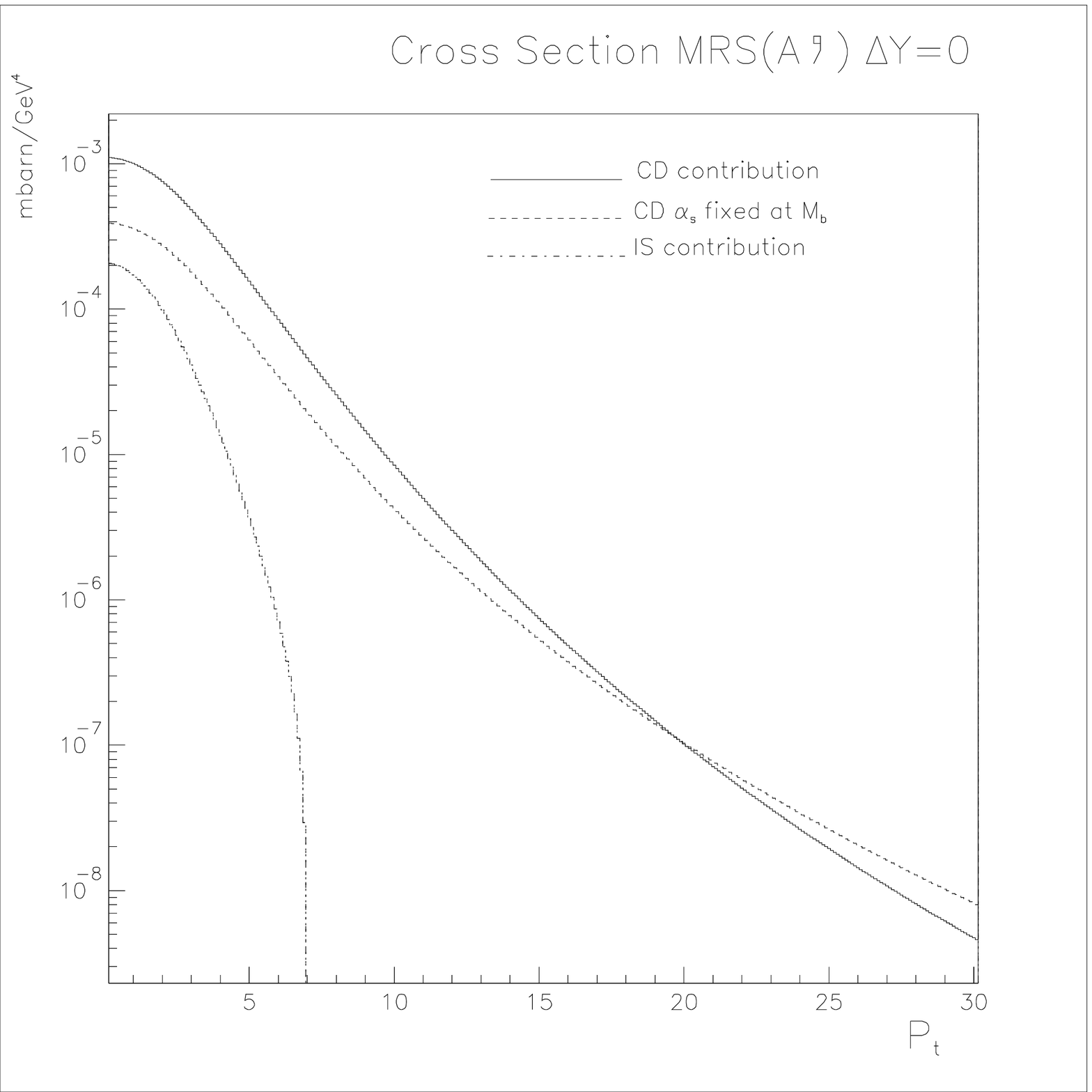,width=3.0in} & \psfig{figure=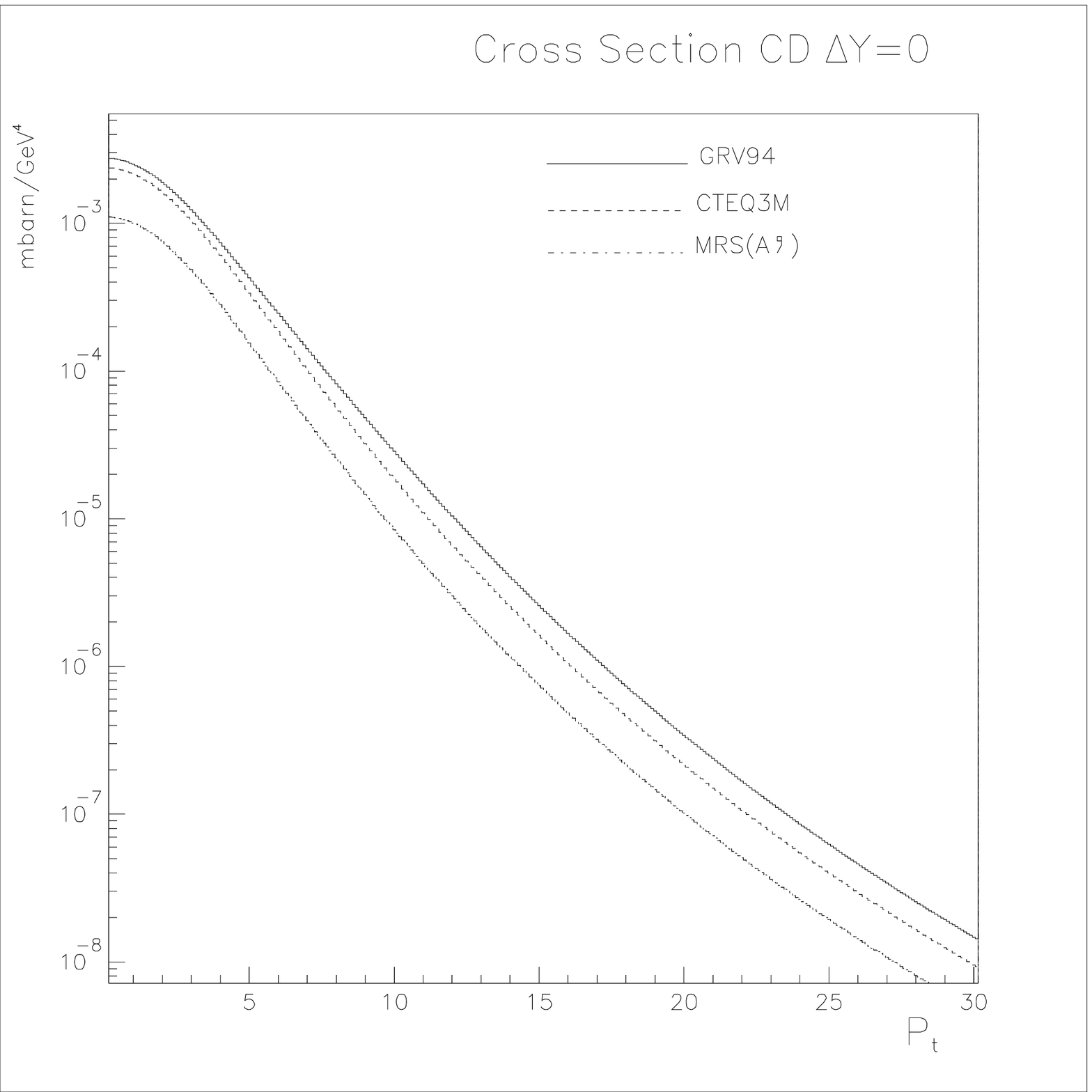,width=3.0in}\\
Fig. \ref{CSECMRS}a  &  Fig.\ref{CSECMRS}b \\
\end{tabular}
\protect\caption{ (a) The cross section for $b \bar b$ diffractive production 
for the coherent diffraction (CD) (Eq. (29)) and the Ingelman - Schlein (IS) 
diffraction ( Eq. (25)); (b) The cross section for $ b \bar b $ for the coherent 
diffraction (CD) in different parameterizations for the gluon structure function.}
\protect\label{CSECMRS}
\end{figure}

\begin{figure}[htbp]
\centerline{\psfig{figure=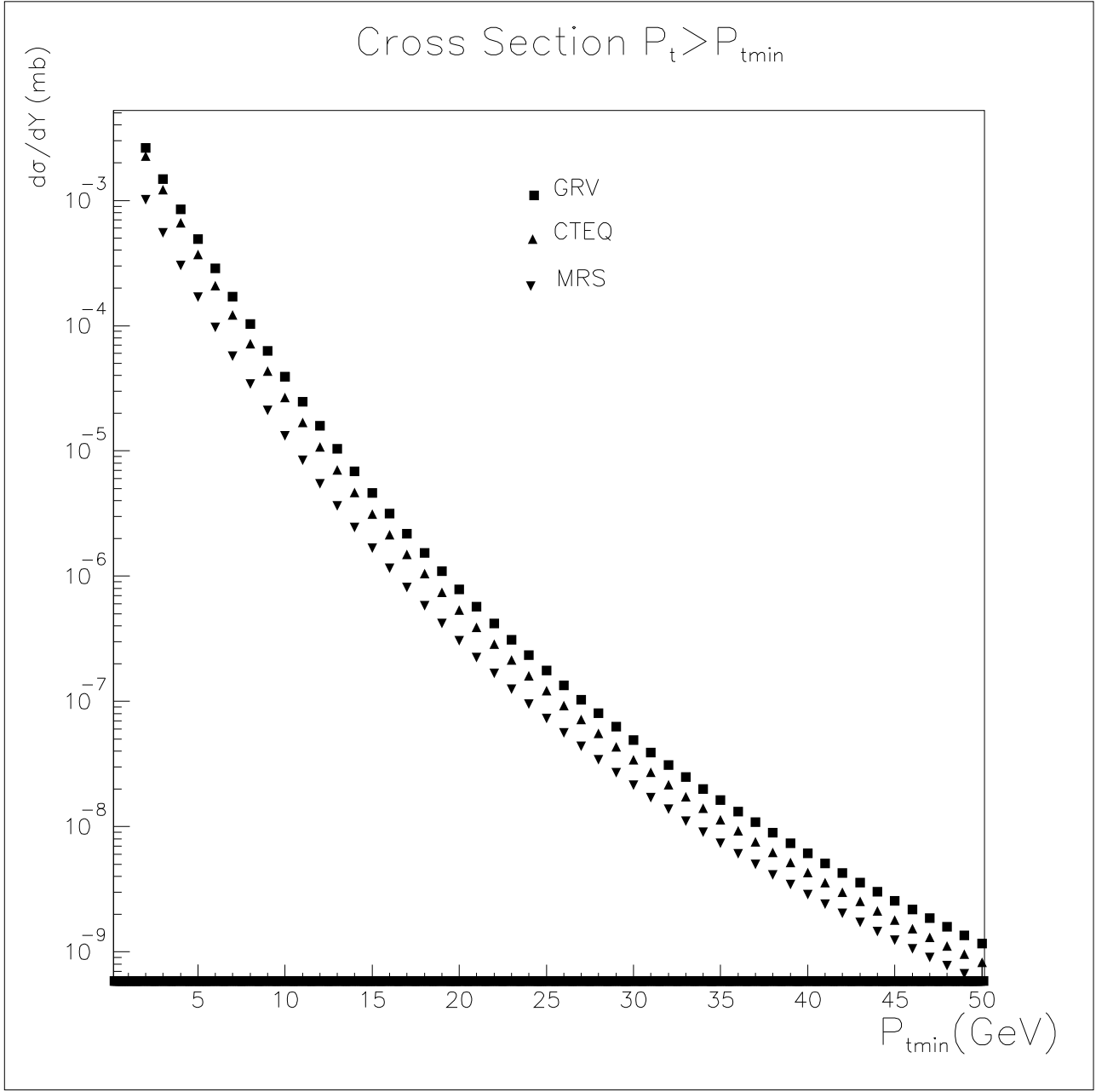,height= 17cm,width=17cm}}
\caption{ The integrated cross section for the coherent diffraction ( Eq.(32)) 
versus $p^{min}_t$ in different parameterizations of the gluon structure 
function. } 
\protect\label{CSECPT}
\end{figure}

The result of a detailed calculation
is presented in Fig. \ref{CSECPT}. To test the sensitivity of our result to
high order QCD corrections we plotted the cross section for the coherent
diffraction for two cases: fixed and running QCD coupling constant. The
difference is rather big but it does not change the main conclusion: the
coherent diffraction gives much bigger cross section than the
Ingelman-Schlein contribution of Eq.(25).  (IS in Fig. \ref{CSECMRS}a). Terefore, the
measurement of the diffractive dissociation in $ b \bar b $ system gives the
unique opportunity to study the CD unlike the deep inelastic processes
where the CD is suppressed.

	One can see from Fig. \ref{CSECMRS}b that the value of the differential
cross section crucialy depends on the parametrization of the gluon structure
 function with the difference about factor 2. This difference encourages us
 to claim
that the measurement of the $b \bar b$ diffractive production could provide
the selection of the parameterization and give an important contribution to
the  extraction of the value of the gluon structure function from experiment.
 
We also calculate the integrated cross section defined as 
\beq \label{29}
\frac{ d \s}{d Y}\,\,=\,\,\int^{\infty}_{p^{min}_t} d p^2_t \int^{+\infty}_{-\infty} 
d \Delta y \int^{\infty}_{0} d q^2_t \,\,\frac{d \s}{ d Y d \Delta y d q^2_t 
d p^2_t}\,\,,
\eeq 
The value of $p^{min}_t$ can be taken from the experimental values 
obtained by the Tevatron Collider experiments. In reference  
Ref. \cite{DZERO}  analysis techniques are used to separate muons
coming from different sources and, in particular, from $b\,\rightarrow 
\,\mu \,+\,\nu \,+\,c $ process for $p^{min}_t \geq$ 5  GeV.
We assumed, in the integration over $q^2_t$, an exponential behaviour of
$d\sigma$ with respect to $q^2_t$

\begin{equation}
\frac {d\sigma}{dq^2_t} = \frac {d\sigma}{dq^2_t}|_{q^2_t=0} e^{-bq^2_t}
\end{equation}

We take  the slope $ b = 4.9 GeV^{-2}$ as it has been measured at
HERA \cite{ZEUS}. In our estimates we took $\as$ = 0.25 which corresponds
to the value of the running QCD coupling constant ($\as(k^2)$ ) at
$k^2 \sim m^2_b$.

Fig. \ref{CSECPT} shows that the value of the integrated cross section is 
not very small and can be meausered by the Tevatron detectors in the next 
run. One can also see that the differences between the estimates in different 
parameterizations is rather big. It is about a factor 2-3 between the highest
values of the cross section in the GRV parameterization and the lowest one  
in the MRS(A') parameterization.

Finally, we would like to stress that the Tevatron provides an unique 
possibility to look inside the microscopic mechanism of diffractive 
dissociation by measuring the coherent diffraction which gives a small 
contribution to the deep inelastic proccesses. The formulae written in 
this paper give us the basis for the Monte Carlo simulation of the 
diffractive events in 3-dimension phase space ($\eta,\phi, p_t$ ) which 
we are going to present in further publications.
This Monte Carlo will provide a more detailed estimates of the experimental 
possibilies at the Tevatron and, we hope, will encourage future experiments 
on large rapidity gap physics. We firmly believe that the 
diffractive dissociation opens a new window to study such difficult 
questions as the Pomeron structure, the matching between hard and soft 
processes and the search of new collective phenomena in QCD related to the 
high density parton system.
	We thank the Fermilab Theory, Computing and Research Divisions for
the hospitality. E.L. is very grateful to LAFEX-CBPF/CNPq  for the 
support and warm atmosphere during his stay in Brazil. We would like to 
thank H.Montgomery for fruitfull discussions in Rio, and to Helio da Motta
for reading the manuscript.


\newpage

\begin{figure}[p]
\centering
\setlength{\unitlength}{1.0in}
\psfig{figure=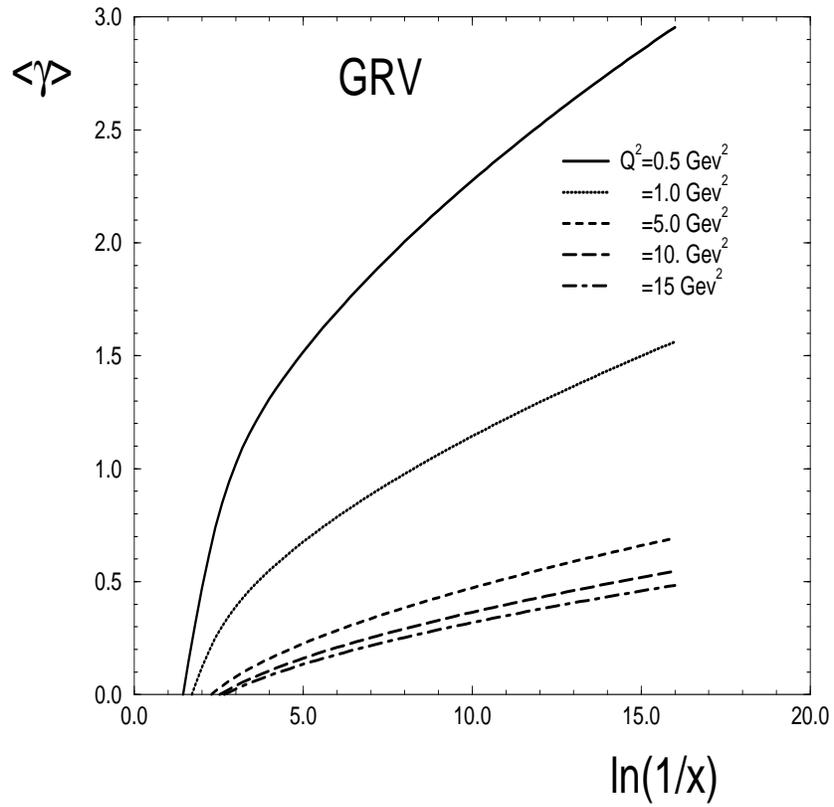,width=6.0in}
\caption{The behaviour of average $< \gamma >$ for the GRV parameterization
 of the gluon structure function.}
\label{GAMMA}
\end{figure}

\end{document}